\documentclass[oneside,a4paper]{article}

\usepackage{hyperref}
\usepackage{changepage}
\usepackage{graphicx}
\usepackage[utf8]{inputenc}
\usepackage[main=english]{babel}
\usepackage{graphicx} 
\usepackage[justification=default]{subfig} 
\usepackage[update]{epstopdf}
\usepackage[labelfont=bf]{caption}
\usepackage{titlesec} 
\usepackage{booktabs}
\graphicspath{{Figures/}}
\usepackage[section]{placeins}
\usepackage{enumerate} 
\usepackage{tablefootnote}
\usepackage{threeparttable}
\usepackage{caption}
\usepackage{subfig}
\usepackage{wrapfig}
\usepackage{cite}
\usepackage{longtable}
\usepackage{multicol}
\usepackage{array}
\usepackage{tabularx}
\usepackage{gensymb}
\usepackage{xcolor}
\usepackage{bbding}
\usepackage{amsfonts}
\usepackage{amsmath,amssymb}
\usepackage{mathptmx}

\usepackage{color}
\usepackage[textwidth = 450pt,top = 80pt, bottom = 60pt]{geometry}
\usepackage{soul}

\usepackage{authblk}
\providecommand{\keywords}[1]{\textbf{\textit{Keywords:}} #1}

\title{Flexible skyline: overview and applicability}
\author{Carlo Bellacoscia}
\affil{Politecnico di Milano\\
Milan, Italy\\
\href{mailto:carlo.bellacoscia@polimi.it}{carlo.bellacoscia@polimi.it} }
\date{}

\begin{document}
\maketitle
\begin{abstract}
Ranking (or top-k) and skyline queries are the most popular approaches used to extract interesting data from large datasets.
The first one is based on a scoring function to evaluate and rank tuples. Its computation is fast, but it is sensitive to the choice of the evaluating function.
Skyline queries are based on the idea of dominance and the result is the set of all non-dominated tuples. This is a very interesting approach, but it can't allow to control the cardinality of the output.
Recent researches discovered more techniques to compensate for these drawbacks. In particular, this paper will focus on the flexible skyline approach.
\end{abstract}


\section{Introduction}
\begin{figure}[h]
        \centering
        \includegraphics[width=0.5\linewidth]{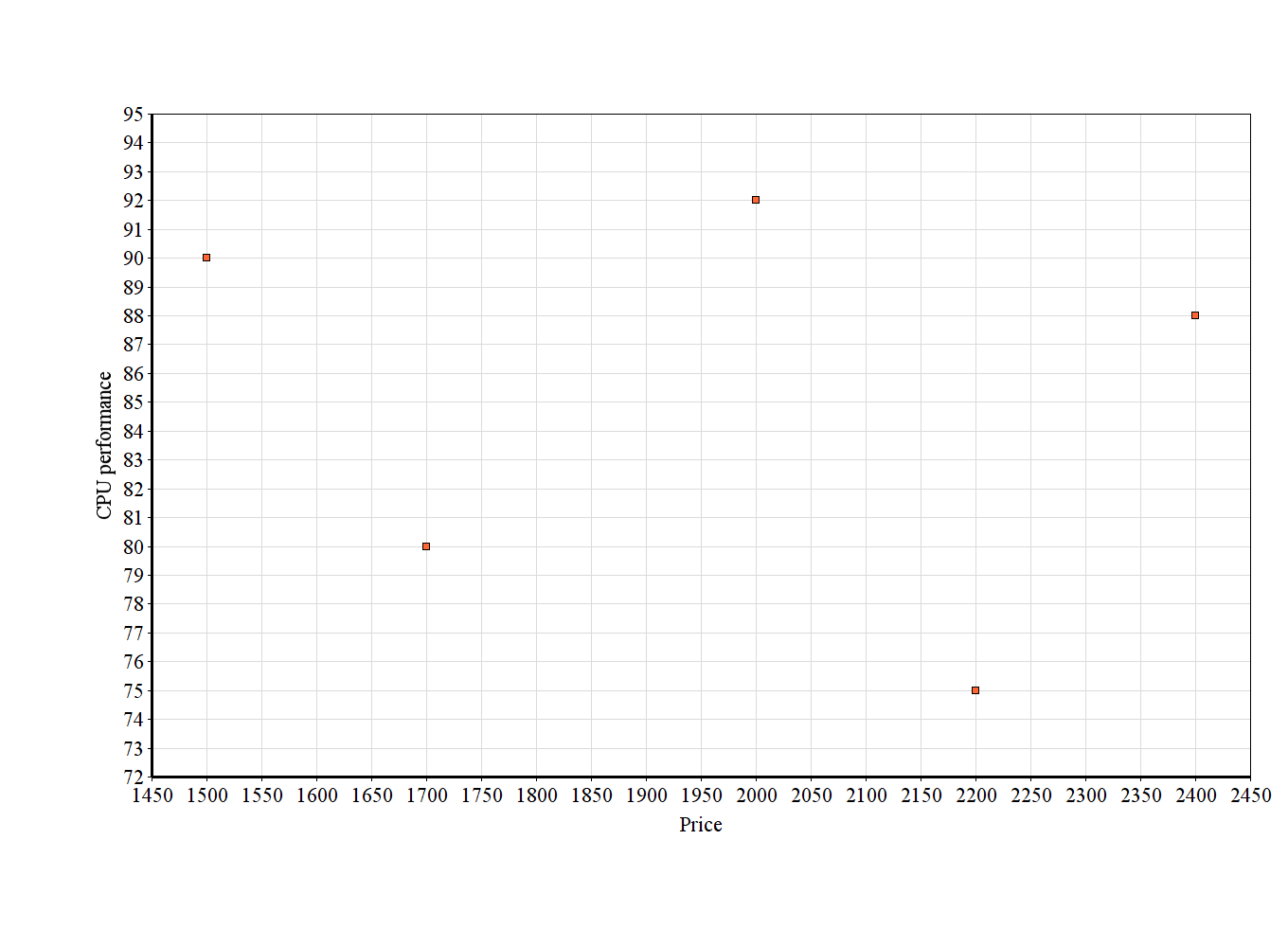}
        \caption{Laptops by price and CPU performance}
        \label{fig:graph}
    \end{figure}

The ranking query approach is a very important concept used in many situations. Nowadays lots of systems work with big databases so most of interactions have the aim to extract only a specific number of data related to a certain set of characteristics.\\
The most used method to do so is the Top-k one. This approach uses a particular function called scoring function that evaluates queries and returns the best k ones.\\
The other method that is widely used in the selection of data from the database is the Skyline one that is based on the idea of dominance; it returns a fixed number of queries that are better than all the others.\\
However, these two methods have some drawbacks. In a multi-objective problem, the first one sum up all the attributes in a single output strictly related to scoring function, so drastic changes in the function modify results.\\
The second one uses the idea of dominance to select the set of data, but doesn't allow to choose the cardinality of the output and modify the preferences about the attributes that have to be used for the evaluation.\\
For example, considering the web search of the best laptop for price and CPU performance, the top-k approach using a scoring function like the sum of all attributes multiplied by corresponding weights would return as output a unique value and the algorithm would generate the list of the best k laptops by both the attributes, price and performance; it wouldn't be possible to do the same research considering a single attribute (price or performance) because the function is fixed. Using the skyline algorithm the output would be the list of laptops that dominate all the others on price and performance, but it wouldn't be possible to choose the number of dominant laptops considered as result and which attribute use for the evaluation.\\
In Table \ref{tab:PCs} and Figure \ref{fig:graph} were reported the data considered in the example and a graphical representation of them.\\
  \begin{table}[h]
        \centering
        \caption{\label{tab:PCs}List of selected PCs}
            \begin{tabular}{lcccccc}
                \toprule
                Brand & CPU performance & Price\\ 
                \midrule
                Dell & 90 & 1500\\
                Apple & 80 & 1700\\
                Microsoft & 92 & 2000\\
                Lenovo & 75 & 2200\\
                Asus & 88 & 2400\\
                \bottomrule
            \end{tabular}
    \end{table}
Alternative approaches were developed to overcome these drawbacks. One of them is the flexible skyline method, that can take into account the different importance of the attributes applying constraints on the weights used in the scoring function, granting a much greater flexibility.\\\\

\section{Classic Approach}
\begin{adjustwidth}{2.5em}{0pt}
    \item \subsection{Top-k}
    The aim of the Top-k algorithm is to take the most significant k tuples from the database according to a function called scoring function.\\
    This function uses attributes of the object and produces a single value output, an example is the sum of each dimension multiplied by a weight that is a number between 0 and 1. The scoring function is the most important part of this approach and in most of the cases this function is monotone; this property provides some features like a faster computation, because if an attribute of an object is linked to the same attribute of another object, the relation will be maintained also in scoring function for every couple of tuples.\\
    
    \begin{math}
        A\; scoring\; function\; f\; is\; a\; function\; f:[0,1]^{d}\to\mathbb{R}^{+}.\; For\; a\; tuple\; t = <v_1,...,v_d>\; over\; \mathbb{R},\; the\\ value\; f(v_1,...,v_d)\; is\; called\; score\; of\; t,\; also\; written\; f(t).\\
        Function\; f\; is\; monotone\; if,\; for\; any\; tuples\; t,s\; over\; \mathbb{R},\; the\; following\; holds:\\
        \forall i\; \in\; {1,...,d}\; t[A_i] <= s[A_i]\; \to\; f(t) <= f(s)\\
        the\; infinite\; set\; of\; all\; monotone\; scoring\; functions\; is\; denoted\; by\; MF.\\
    \end{math}
    In multi-objective problem this method returns a single value that is fixed by scoring function and it's editable only in the weights of the function. This grants some flexibility, but it isn't enough to be suitable to every case.\\
    \begin{figure}[h]
        \centering
        \includegraphics[width=0.5\linewidth]{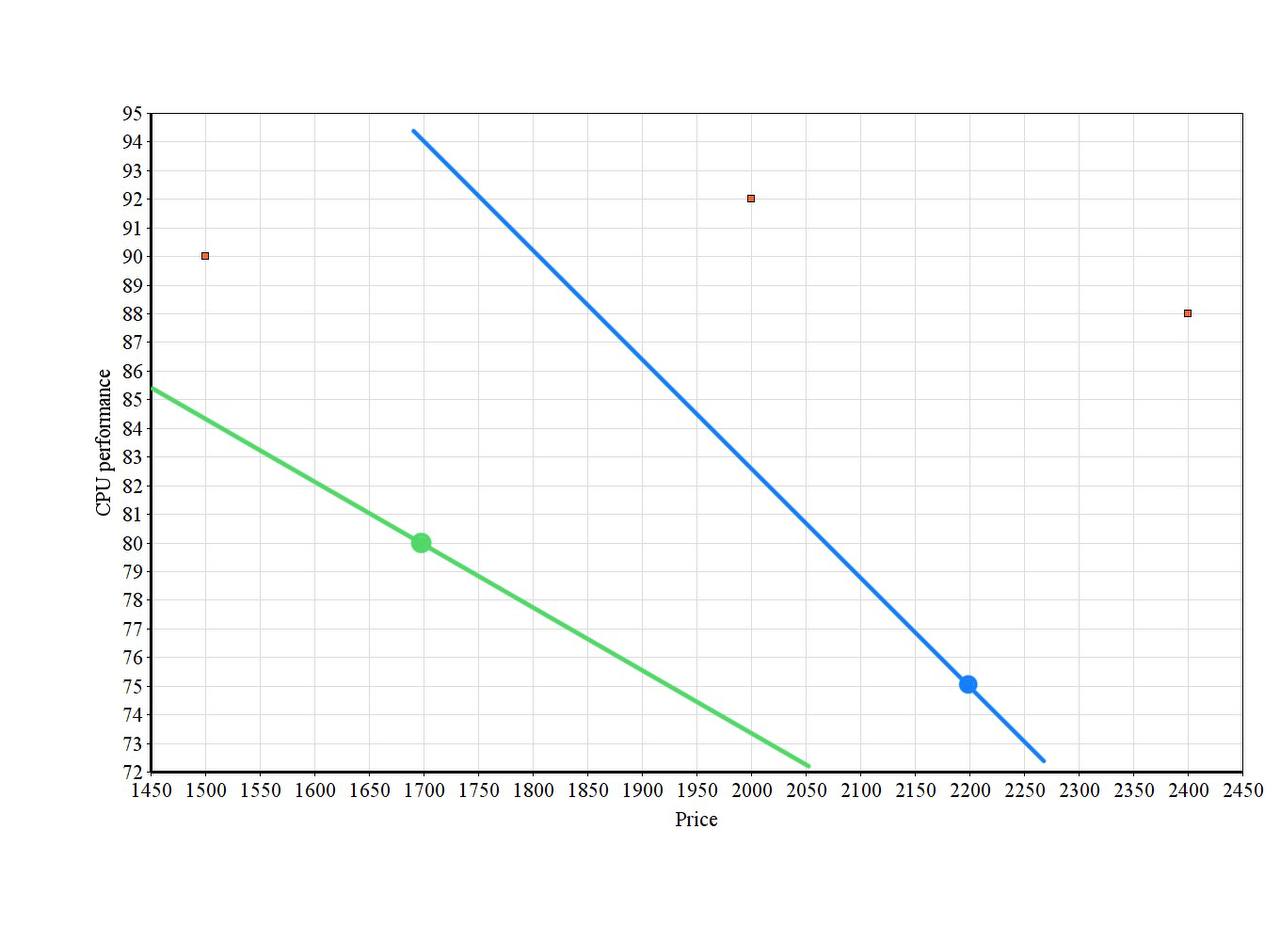}
        \caption{Ranking of laptops by two different scoring functions}
        \label{fig:top-k}
    \end{figure}
    The Figure \ref{fig:top-k} highlights the difference of scoring function in the example explained in the introduction, where the two lines, the green one and the blue one select completely different data because the weights in them change. Then the user should know a priori how much a feature of the tuple is important in his search.\\
    The importance of scoring function was highlighted in \cite{top-k}, where the top-k has been tested in strict conditions and the performance varied significantly depending on used function.\\
    The possibility to reduce this lack was proposed in \cite{uncertain}, where was introduced the notion of uncertainty in the selection of the weights. The algorithm proposed uses a preference region where the scoring function can act.\\ 
    In \cite{web} the applicability of top-k in access of web-databases was discussed, where data are sparse and with different features. Then two extensions of existing algorithms were presented, $Upper$ and $Pick$. The first one stores the tuples that have the best chance to be the optimal, the second one minimizes a function that represents the "distance" between the current execution and the final state. These operators improve performance but they were tested assuming that there is only one list of objects (one S-Source), sorted by descending scores, and multiple random access to a single attributes (multiple R-Source).\\
    \item \subsection{Skyline}
    This approach is based on the idea of pareto dominance.\\
    
    \begin{math}
    Let\; A\; be\; the\; set\; of\; relevant\; attributes\; a\; tuple\; has\; A=[a_1,...,a_d],\; r\; be\; the\; set\; of\; tuples\; that\; are\\ considered\; and\; let\; t,s\; be\; two\; tuples\;\in\; r:\\
    t\prec s,\; iff \forall i,\; 1<=i<=d\; \to\; t[A_i]<=s[A_i],\; and\; \exists j,\; 1<=j<=d\; \wedge t[A_j]<s[A_j]\\
    and\; so,\; the\; formal\; definition\; of\; Skyline\; over\; r\; is:\\
    Sky(r)=\{t\; \in r\; |\; \nexists s\; \in r,s\; \prec\; t\}
    \end{math}\\
    
    The skyline queries compute the dominance for each tuple and return the ones that aren't dominated by others. This method does not suffer from the same drawbacks of top-k because it isn't connected to a scoring function, but it hasn't the flexibility of choosing weights, so the output can't be controlled in terms of size and way to use attributes.\\
    In this case the approach requests to the user to select the tuples that he needs after the computation.\\
    \begin{figure}[h]
    \centering
    \includegraphics[width=0.5\linewidth]{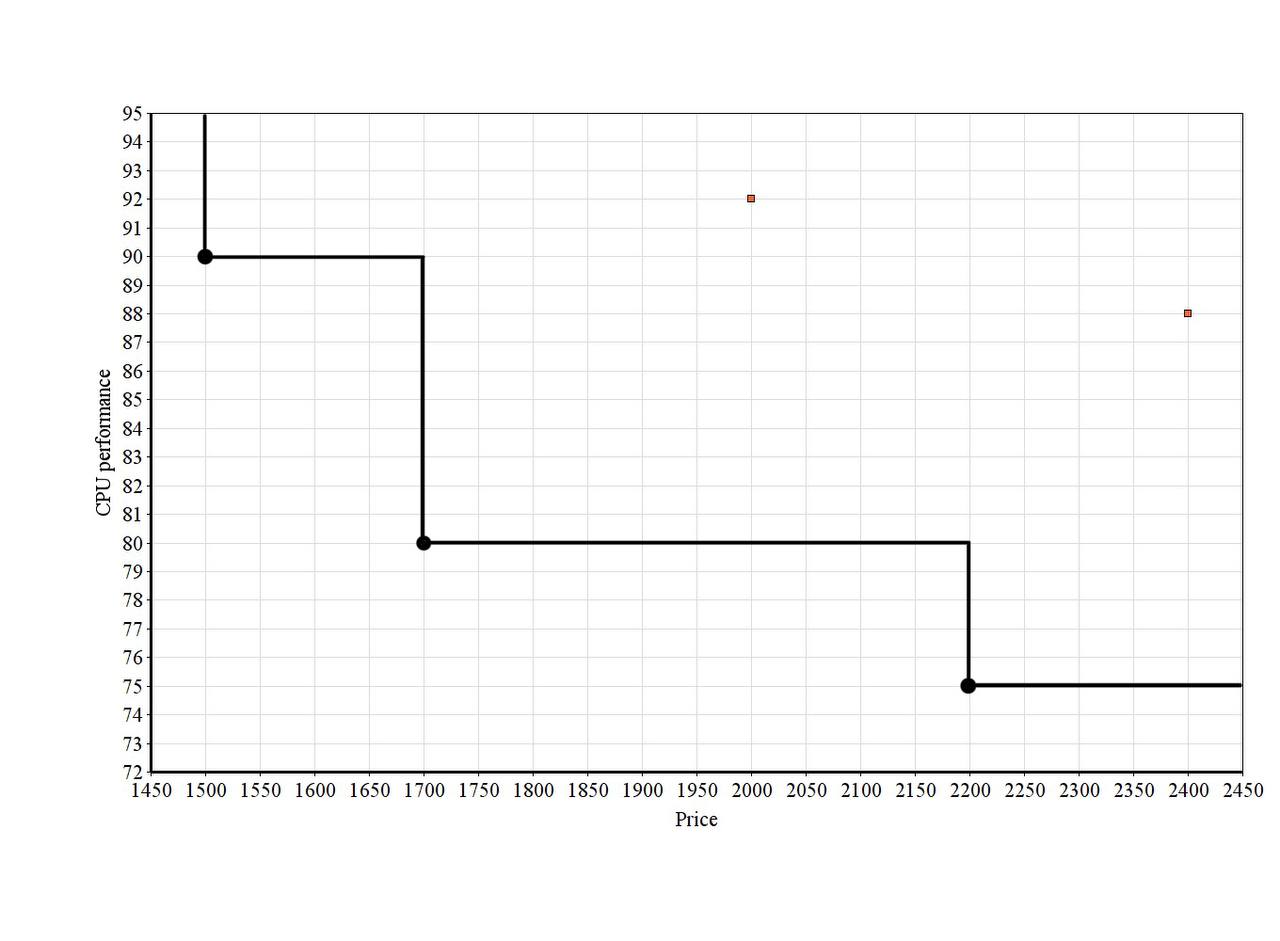}
    \caption{Skyline of laptops}
    \label{fig:skyline}
    \end{figure}
    In Figure \ref{fig:skyline} there is the example of the Skyline described in the introduction, where the algorithm offers three different laptop that aren't dominated by others, but if the user needs only one preference he has to select one of them. This intrinsic property of Skyline encouraged in making new approaches that fill this lack, in the \cite{overcome} some approaches were proposed in order to try to control the cardinality of the output.
\end{adjustwidth}

\section{Alternative Approach}
Alternative approaches were defined to overcome the classical approach drawbacks.\\
The first problem that has been faced was size tuning of Skyline, in \cite{size, flexible} this problem was handled, but in these works it was highlighted that this approach isn't so effective in the resolution because converts complex problems in simple ones and this could produce tie and make the algorithm expensive in time consuming.\\
Some of the alternative approaches will be described in the following paragraphs.
\begin{adjustwidth}{2.5em}{0pt}
    \item \subsection{Fagin's Algorithms}
    Fagin's work was the first one to face the problem of multi-objective top-k queries. In \cite{fa1} he reached a solution to combine information from different subsystems in a natural way.\\
    An algorithm that combines information with the notion of upper and lower bound was presented.\\
    In \cite{fa2} the aim was to expand classic algorithm for ranking queries to use multimedia databases with objects that have inherently fuzzy attributes.\\
    He proposed two algorithms:
    \begin{itemize}
        \item FA\\
        This algorithm considers two types of access to data: the sorted one (or sequential), in which the schedule is managed by a list of objects; the second one is the random access, where the system enters into a specific field of an object in the list.\\
        FA deals with fuzzy data using an aggregation function; assuming that this aggregation is independent, every correct algorithm, with high probability, incurs a similar middleware cost in the worst case.\\
        The algorithm is made of the following steps:
        \begin{enumerate}
            \item sorted access of m lists \begin{math}L_i\end{math} until there are at least k "matches", then wait to find the k matches in other lists.
            \item for every object do a random access to  \begin{math}L_i\end{math} to find the \begin{math}i-th\end{math} field.
            \item compute the grade and populate a list of the best ones.\\
        \end{enumerate}
        \item TA\\
        The TA algorithm is made of the following steps:
        \begin{enumerate}
            \item sorted access of m lists \begin{math}L_i\end{math}. As an object is seen in sorted access, do a random access to find the grade \begin{math}x_i\end{math} in every list \begin{math}L_i\end{math}. Then compute the grade, t(R) for R object. If the grade is one of the k best remember object and grade.
            \item for every list compute a \begin{math}threshold\; value\; \tau \end{math} to be \begin{math}t(x_1,...,x_m)\end{math}. As soon as at least k object have been seen whose grade is at least equal to \begin{math} \tau \end{math}, then halt.
            \item let Y be the resulting list. As soon as Y has at least k object then it gives the output.
        \end{enumerate}
    \end{itemize}
    In \cite{restr} the FSA algorithm was presented, that combines the TA and FA. This is a framework for processing multisource top-k queries defined by scoring functions in which weights are only partially specified, thus providing the required flexibility that is necessary in many application developed nowadays. This features are obtained by using the notion of \begin{math}F-Dominance\end{math}:\\
    \begin{math}
    Let\; be\; F\; a\; set\; of\; monotone\; scoring\; functions.\; A\; tuple\; t\; F-Dominates\; another\; tuple\; s\; \neq\; t,\; denoted\; by\; t\; \prec_F\; s,\; if\; \forall f\; \in\; F.\; f(t) <= f(s).\\
    \end{math}
    \item \subsection{Epsilon-Skyline }
    In \cite{eps} a new approach that solves the problem of the fixed cardinality of Skyline was described. The $\varepsilon$ operator gives the possibility to control the size of the query in shrinking or expanding direction.\\
    Assuming that each dimension of the data is normalized between [0,1], the algorithm is based on $\varepsilon-Dominance$:\\
    \begin{math}
        Given\; a\; set\; of\; d-dimensional\; object\; T\; with\; weights\; W\; =\; {w_i\; |\; i \in [1,d],\; 0\; <\; w_i\; <=\; 1},\; and\; a\\ constant\; \varepsilon\; \in\; [-1,1],\; for\; any\; two\; objects\; t_1\; =\;{t_1[1],...,t_1[d]}\; and\; t_2\; =\; {t_2[1],...,t_2[d]},\; if\; \forall i\; \in\; [1,d].\; t_1[i]*w_i\; <=\; t_2[i]*w_i\; +\; \varepsilon,\; and\; \exists j\; \in\; [i,d],\; t_1[j]\; <=\; t_2[j],\; then\; t_1\; is\; said\; to\; \varepsilon -dominate\; t_2,\; denoted\; as\; t_1\; \prec_\varepsilon\; t_2.\; The\; \varepsilon -skyline\; of\; the\; dataset\; is\; the\; set\; of\; all\; objects\; in\; T\; that\\ are\; not\; \varepsilon -dominated\; by\; any\; other\; objects.
    \end{math}\\
    \begin{figure}[h]
        \centering
        \includegraphics[width=0.5\linewidth]{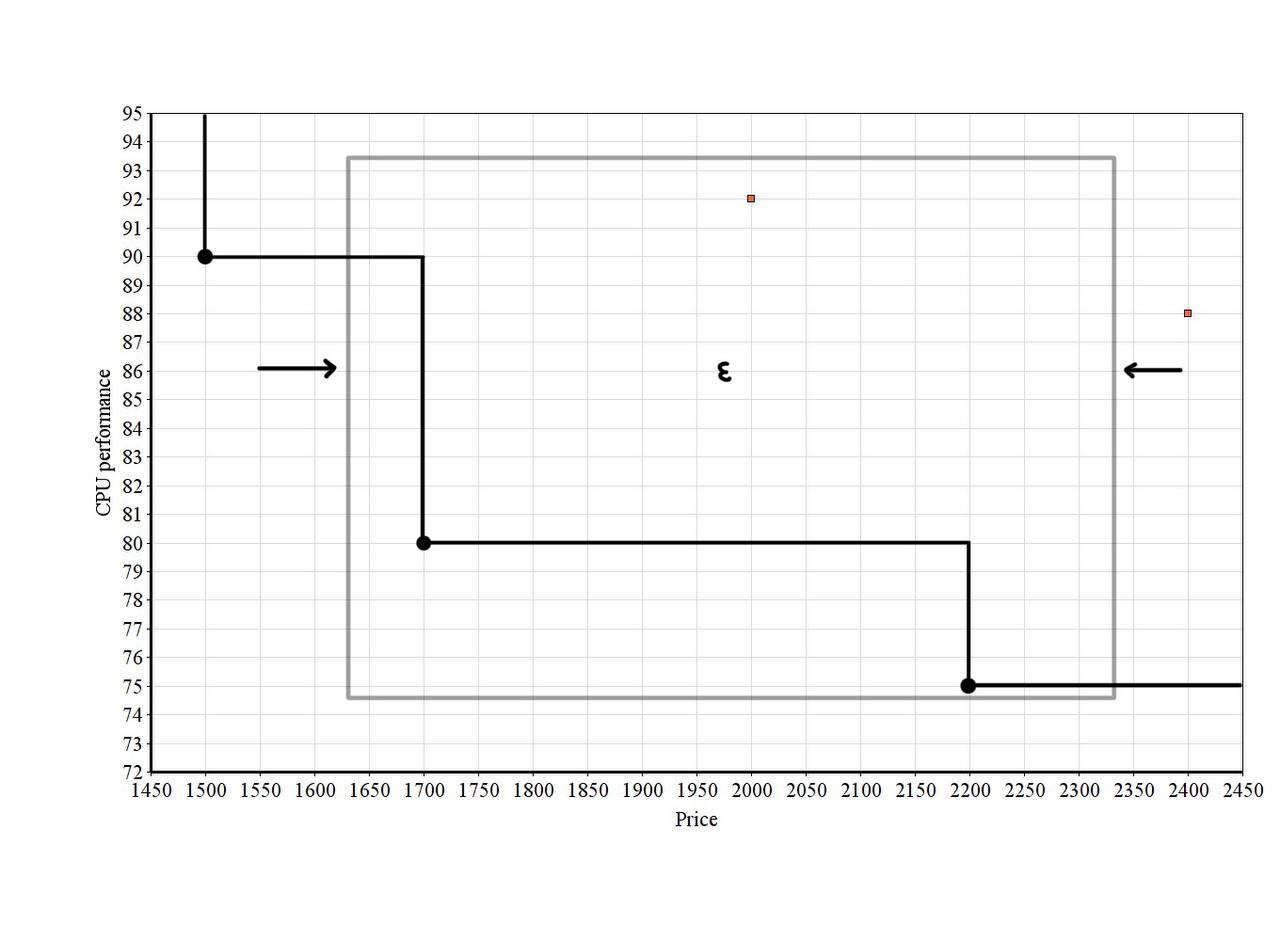}
        \caption{$\epsilon-Skyline$ of laptops}
        \label{fig:eps}
    \end{figure}
    Using this method for the proposed example, defining the width of the $\epsilon$ region it's possible to manage the cardinality of the data selected by the query and the preferences of the user.\\
    In Figure \ref{fig:eps} it's clear how this approach restricts the output choosing the best combination of attributes, giving as result only two of the three tuples that the computation of the pareto dominance would return.\\
    \item \subsection{ORD and ORU}
    The paper \cite{ord} deals with the problems in personalization, controllable size of the output and flexibility in preference specification, with two operators ORD and ORU.\\
    Given the seed vector w and the required output size m, ORD reports the records that are \begin{math} \rho-dominated \end{math} by fewer than k others, for the minimum \begin{math} \rho \end{math} that produces exactly m records in the output.\\
    Given the seed vector w and the required output size m, ORU reports the records that belong to the top-k result for at least one preference vector within distance \begin{math} \rho \end{math} from w, for the minimum \begin{math} \rho \end{math} that produces exactly m records in the output.\\
    This is a practical and scalable performance that could be useful in sparse or skewed datasets.\\
    With these two operators it's possible to manage the dimension of the output using two main ideas: inflection radius and convex hull. Starting from the weights vector, the ORD computes $\rho$ to obtain the minimum inflection radius that contains the exact number m of dominated tuples, with m equals to the desired output.\\
    The ORU uses the idea of convex hull, that is the smallest polytope that encloses all the tuples of the dataset; this figure is created by the calculation of the upper hull, that is the set of all the facets of the convex hull where the normal vector is directed towards the positive section of the plane. The top region of t is calculated as the region where t has the highest score relative to the weights vector. The ORU recursively computes the top region, first checking the adjacent tuples then, if nothing is found, expand the upper hull. Once the best set of tuples is found, the output will be the defined number m of tuples with the normal vector with the minimum distance $\rho$ from the weights vector.\\
\end{adjustwidth}

\section{Flexible Skyline}
After the alternative approaches described above, in \cite{start, restr, mix} the idea of Flexible-Skyline was introduced that overcomes the problems of top-k and skyline queries.\\
Algorithms used in this approach are based on the notion of \begin{math} F-Dominance \end{math} already mentioned in "Fagin's Algorithms" section. Two operators were introduced: ND, non dominated tuples, and PO, potentially optimal ones.\\
Flexible Skyline operators behavior is the same as Skyline ones, but applied to a limited set of monotone scoring function \begin{math} F\; \subseteq\; MF \end{math}.
\begin{adjustwidth}{2.5em}{0pt}
    \item \subsection{ND computation}
    In this section there is an overview of non dominated tuples computation, ND(r;$F$), where $F$ is a set of MLW (monotonically transformed, linear-in-the-weights).\\
    The first approach to compute ND uses $F-Dominance \; Test$:\\
    \begin{math}
        Let\; F\; be\; a\; set\; of\; MLW\; functions\; subject\; to\; a\; set\; C=\{C_1,...,C_c\}\; of\; linear\; constraints\; on\; weights,\\ where\; C_j\; =\; \sum_{i=1}^{d}\; a_{ji}w_i\; <=\; k_j\; (for\; j\; \in\; {1,...,c}).\; Then,\; t\; \prec_F\; s\; iff\; the\; following\; linear\\ programming\; (LP)\; problem\; in\; the\; variables\; W\; =\; (w_1,...,w_d)\; has\; a\; non\; negative\; solution:\\
        minimize:\qquad \Lambda \cdot \sum_{i=1}^{d} w_i(g_i(s[A_i]-g_i(t[A_i])),\\
        subject\; to:\\
        w_i \in [0,1] \quad i \in \{1,...,d\}\\
        \sum_{i=1}^{d} w_i = 1,\\
        \sum_{i=1}^{d} a_ji w_i <= k_j \quad j \in \{1,...,c\}
    \end{math}.\\\\
    This method is likely to be time-consuming because the LP problem needs to be solved for each object.\\
    An alternative is to compute the $F-Dominance$ regions of tuples and not consider the ones that belong to at least one of these regions.\\ 
    F -dominance Region:\\
    \begin{math}
        Let\; F\; be\; a\; set\; of\; MLW\; functions\; subject\; to\; a\; set\; C=\{C_1,...,C_c\}\; of\; linear\; constraints\; on\; weights,\\ where\; C_j\; =\; \sum_{i=1}^{d}\; a_{ji}w_i\; <=\; k_j\; (for\; j\; \in\; {1,...,c}).\; Let\; W^{(1)},...,W^{(q)}\; be\; the\; vertices\; of\; W(C).\\ The\; dominance\; region\; DR(t;F)\; of\; a\; tuple\; t\; under\; F\; is\; the\; locus\; of\; points\; in\; [0,1]^d\; defined\; by\; the\\ q\; inequalities:\\
        \Lambda \cdot \sum_{i=1}^{d}\; w_i^{(l)} g_i(s[A_i])\; >=\; \Lambda \cdot \sum_{i=1}^{d}\; w_i^{(l)} g_i(t[A_i]), \;\;\;\;\; l \in \{1,...,q\}
    \end{math}.\\\\
    The only significant overhead introduced by this approach is the enumeration of the vertices of W(C), that has to be done just once.\\\\
    The notation and then the list of used algorithms are described below.\\
    The 'S' or the 'U' at the beginning of the name means that the algorithm is Sorted alternative, in which the data have a pre-processing of sorting by F-Dominance, or Unsorted.\\
    Another distinction is between algorithms with LP in the name, that use Linear Problem to compute the dominance, and the ones with VE in the name that make this control with the idea of vertex enumeration.\\
    Then there is a difference between 1-phase algorithm, where the F-dominance is computed directly, and 2-phases one, in which the Sky operator is used before the F-dominance (the first one has 1 in the name and the second has 2).\\
    The last type of algorithm interleaves dominance (Sky) and F-dominance and it is represented by an 'F' in the name.\\
    The algorithms are:\\
    SVE1, SVE1F, SVE2, SLP2, UVE2, ULP2.
    \item \subsection{PO computation}
    Even in the computation of the potential optimal tuples there are some different algorithms.\\
    First it's useful an overview of the notation. The "PO" at the beginning means potential optimal. Then the "P" or "D" represent the type of test used, the first one for "Primal PO Test", the second one for "Dual PO Test".\\
    Primal PO Test was explained by this theorem:\\
    \begin{math}
        Let\; F\; be\; a\; set\; of\; MLW\; functions\; subject\; to\; a\; set\; C=\{C_1,...,C_c\}\; of\; linear\; constraints\; on\; weights,\\ where\; C_j\; =\; \sum_{i=1}^{d}\; a_{ji}w_i\; <=\; k_j\; (for\; j\; \in\; {1,...,c}),\; and\; such\; that\; h\; is\; strictly\; monotone.\; Let\\ ND(r;F)=\{t_1,...,t_{\sigma},t\}.\; Then,\; t\; \in\; PO(r;F)\; iff\; the\; following\; LP\; problem\; in\; the\; variables\; W=(w_1,...,w_d)\; and\; \phi\; has\; a\; strictly\; positive\; optimal\; solution:\\
        maximize:\qquad \phi,\\
        subject\; to:\\
        \Lambda \cdot \sum_{i=1}^{d} w_i(g_i(t[A_i]-g_i(t_j[A_i])) + \phi \quad j \in \{1,...,\sigma\},\\
        \sum_{i=1}^{d} a_ji w_i <= k_j \quad j \in \{1,...,c\},\\
        w_i \in [0,1] \quad i \in \{1,...,d\}\\
        \sum_{i=1}^{d} w_i = 1,\\
    \end{math}\\
    The Dual PO Test is based on the notion of convex combination of a set of tuples, by the marginal scores of them. Convex combination is defined as follow:\\
    \begin{math}
        Given\; tuples\; t_1,...,t_n,\; n>1,\; and\; monotone\; transform\; g_i,\; 1\; <=\; i\; <=\; d,\; a\; tuple\; s\; is\; a\; convex\\ combination\; (through\; the\; g_i's)\; of\; (the\; marginal\; scores\; of)\; t_1,...,t_n\; if\; there\; exists\; \alpha_1,...,\alpha_n\; such\\ that\; \alpha_j\; \in\; [0,1]\; for\; 1\; <=\; j\; <=\; n,\; \sum_{j=1}^{n}\; \alpha_j\; =\; 1,\; and\\ g_i(s[A_i])\; =\; \sum_{j=1}^{n}\; \alpha_j g_i(t_j[A_i]) \quad i \in \{1,...,d\}.
    \end{math}\\\\
    Now Dual PO Test can be defined:\\
    \begin{math}
        Let\; F\; be\; a\; set\; of\; MLW\; functions\; subject\; to\; a\; set\; C=\{C_1,...,C_c\}\; of\; linear\; constraints\; on\; weights,\\ where\; C_j\; =\; \sum_{i=1}^{d}\; a_{ji}w_i\; <=\; k_j\; (for\; j\; \in\; {1,...,c}).\; Let\; W^{(1)},...,W^{(q)}\; be\; the\; vertices\; of\; W(C)\; and\\ let\; ND(r;F)\; =\; \{t_1,...,t_{\sigma},t\}.\; Then\; t\; \in\; PO(r;F)\; iff\; there\; is\; no\; convex\; combination\; s\; of\; t_1,...,t_{\sigma}\\ such\; that\; s\; \prec_F\; t,\; i.e.,\; iff\; the\; following\; linear\; system\; in\; the\; variables\; \alpha\; =\; (\alpha_1,...,\alpha_{\sigma})\; is\\ unsatifiable:\\
        \Lambda \cdot \sum_{i=1}^{d}\; w_i^{(l)} (\sum_{j=1}^{\sigma}\; \alpha_j g_i(t_j[A_i])) <= \Lambda \cdot \sum_{i=1}^{d}\; w_i^{(l)}g_i(t[A_i]) \quad l \in \{1,...,q\},\\
        \alpha_j\; \in\; [0,1] \quad j \in \{1,..,\sigma\},\\
        \sum_{j=1}^{\sigma}\; \alpha_j\; =\; 1.
    \end{math}\\\\
    Testing if tuple t is potentially optimal respect to $\sigma$ other tuples could be very time expensive if $\sigma$ is large. There is a variation of the algorithm that increments $\sigma$ at each step. Then the "I" means that was used an incremental process, the "F" means the full size of $\sigma$.\\
    As in ND computation, there is a choice between "one-phase" ("1" in the name), in which the algorithm discards non-PO(r;$F$) tuples starting from original relation r (non-ND-based), and "two-phase", where first the algorithm computes ND(r;$F$) and then removes the non-PO(r;$F$) (ND-based).\\
    Then the algorithms are:\\
    POPF1, PODF1, POPF2, PODF2, POPI1, PODI1, POPI2, PODI2.\\
\end{adjustwidth}

\section{Results}
In \cite{start} some experiments that highlights the results of Flexible-Skyline approach were presented.
\begin{adjustwidth}{2.5em}{0pt}
    \item \subsection{Efficiency}
    Using real databases, the computation of ND shows that sorting the input and vertex enumeration is a better alternative. The complexities for each algorithm are listed in Table \ref{tab:complexity}, where:
    \begin{itemize}
        \item $lp(x,y)$ is the resolution of linear problem
        \item $ve(c)$ is the complexity of vertex enumeration
        \item c is the number of costraints
        \item d is the number of variables
        \item q is the number of vertices.
        \item $C_{ND}$ is the complexity of ND
    \end{itemize}
    \begin{table}[h]%
    \centering
    \caption{\label{tab:complexity}Complexity registered in \cite{start}}
        \subfloat[][Computating ND]{
        \begin{tabular}{lccccc}
            \toprule
            \toprule
            Algorithm & First phase & Second phase\\
            \midrule
            ULP2 & $O(N^2)$ & $O(|SKY|^2 \cdot lp(c,d))$ \\
            UVE2 & $O(N^2)$ & $O(ve(c) + |SKY|^2 \cdot q)$ \\
            SLP2 & $O(N \cdot (log\;N + |SKY|))$ & $O(|SKY|\cdot |ND| \cdot lp(c,d))$ \\
            SVE2 & $O(N \cdot (log\;N + |SKY|))$ & $O(ve(c) + |SKY|\cdot |ND| \cdot lp(c,d))$ \\
            SLP1, SVE1F & $O(ve(c) + N \cdot (log\;N + |ND| \cdot q)$ & \\
            \bottomrule
            \bottomrule
            \label{Conical}
        \end{tabular}
        } \quad
        \subfloat[][Computating PO]{
        \begin{tabular}{lcccccr}
            \toprule
            \toprule
            Algorithm & Complexity\\
            \midrule
             POPF2 & $C_{ND} + O(|ND| \cdot lp(|ND| + c,d))$\\
             PODF2 & $C_{ND} + O(|ND| \cdot lp(q,|ND|))$\\
             PODI1 & $O(N \cdot log\;N \cdot lp(q,N))$ &\\
             POPI2 & $C_{ND} + O(|ND| \cdot log\;|ND| \cdot lp(|ND| + c,d))$\\
             PODI2 & $C_{ND} + O(|ND| \cdot log\;|ND| \cdot lp(q,|ND|))$\\
            \bottomrule
            \bottomrule
            \label{RAO}
        \end{tabular}
        }
    \end{table}
    Some of the algorithms used are sensitive on the database, when it becomes challenging one of the best alternative is SVE1, but SVE2 prevails if it isn't so complex. If more selective constraints are considered, like in SVE1F, the algorithm leads better performance, because more constraints means less tuples to check.\\
    In computing PO it's preferable to use ND-based algorithms (the opposite ones require a large number of PO tests) and Dual PO Test, because they give an advantage in performance. In the case of complex databases incremental approach is better than the variant because the full-size $\sigma$ is time expensive. PODI2 is the best alternative, unless ND value is small, where PODF2 prevails.\\
    Overall the experiments showed that F-skyline operators can be used instead of standard skyline, especially considering the flexibility that provides.

    \item \subsection{Variant}
    Considering an environment with m $>=$ 1 processors with shared memory the computation of ND and PO could be done in parallel. First the dataset, $r$ has to be divided in m partitions $r_i$. This algorithm follow these steps:
    \begin{enumerate}
        \item determine the so called "local" ND($r_i;F$)'s (or PO($r_i;F$) ), $\forall i \in \{1,...,m\}$.
        \item apply ND (or PO) on the union of the local ones to obtain the "final" one.
    \end{enumerate}
    \item \subsection{Comparison}
    \begin{table}[h]
    \centering
    \caption{\label{tab:comparison}Comparison between considered approaches}
        \begin{tabular}{lcccccc}
            \toprule
            \toprule
            & Cardinality Control & Personalization & No assumption\\ 
            \midrule
            Top-k & \Checkmark & \Checkmark & -\\
            Skyline & - & - & \Checkmark\\
            FA/TA & \Checkmark & - & \Checkmark\\
            $\epsilon$-Skyline & \Checkmark & \Checkmark & \Checkmark\\
            ORD/ORU & \Checkmark & \Checkmark & \Checkmark\\
            Flexible Skyline & \Checkmark & \Checkmark & \Checkmark\\
            \bottomrule
            \bottomrule
        \end{tabular}
    \end{table}
    ("No Assumption" means that the approach doesn't need some specific assumption on the dataset).\\\\
    In this paper some alternative approaches were described to overcome the main drawbacks of classical ones in selecting the best results in a search.
    The concepts that have been explained in this paper was summarized in Table \ref{tab:comparison}.\\
    FA and TA are strictly sensitive to the random access, $\epsilon$-Skyline and ORD and ORU become time expensive with complex databases, but they give their best in sparse datasets.\\
    Focusing on Flexible Skyline and his comparison with classical approaches, top-k and Sky, F-Skyline are more interesting in results because they overcome completely the drawbacks described in first section but they introduce lots of complexity computation to reach the result, so top-k queries take a fraction of the time needed also from the fastest algorithm possible (SVE1F); by the way, the F-Skyline results are much more useful.\\
    The most important advantage respect to top-k is that the flexible approach reaches the same flexibility of ranking queries with the reduction of the space of the data with some additional constraints in the set of monotone scoring functions. Using different values of constraints, it's possible to modify the result in the number of the output tuples and in the selection of the attributes reducing the area of operability. This idea requires less knowledge from the user to select a personalized result.\\
    The F-Skyline could be preferred to the classical Skyline, that has an interesting outcome but it is not controllable in any way. Even if with the Flexible approach it's not possible to define the precise cardinality of the output, its flexibility makes the users able to control it by the constraints.\\
    In terms of performances, F-Skyline requires the same time of Sky in two-phases approach, with SVE2, but if it's possible to use SVE1F, the flexible approach overcomes the standard one.\\
    Using F-Skyline to the proposed example, the user could define a constraint for the perfomance attribute, that could return the top left tuple instead of the set of pareto dominant ones, as shown in Figure\ref{fig:skyline}.\\
    The computation of the F-Skyline is around few seconds so this method could be used in all the situation of the classical approach, but it could be very interesting also in other areas. In \cite{selection} the concept of $F-Dominance$ was used to do feature selection in a Machine Learning problem, but the flexibility that this approach leads can be used also in user preference in social network or for data mining in an e-commerce. 
\end{adjustwidth}

\section{Conclusion}
The aim of the paper was to describe the process to surpass limitations of classical approach. Some approaches were presented and the attention was focused on Flexible-Skyline.\\
This method adds complexity in computation but experimental results show that this does not increase time required so much.\\
The Flexible Skyline approach could be the best choice if it's required a personalization by the user without knowing the weight of the attributes. Top-k queries need more precision in setting the desired weights but, in case of the need of fast computation, this method is preferred.
Compared to the Skyline, the F-Skyline has almost the same computation time so in situations where the Skyline approach is usable, the Flexible one adds flexibility without drawbacks.\\
This features make the Flexible-Skyline a very useful approach that could solve the multi-objective problem.
\bibliographystyle{plain}
\bibliography{refs.bib}
\end{document}